\UseRawInputEncoding

\documentclass[ reprint,  amsmath,amssymb,  aps,prb]{revtex4-1}%
\usepackage[dvipdfmx]{graphicx}
\usepackage{dcolumn}
\usepackage{bm}
\usepackage{amsmath}
\usepackage{amsfonts}
\usepackage{amssymb}%
\usepackage{color}
\usepackage{autobreak}
\providecommand{\U}[1]{\protect\rule{.1in}{.1in}}

\begin{document}

\begin{abstract}

\end{abstract}

%

\preprint{APS/123-QED}%
%

\title{Dynamics governed by symmetry-protected exceptional rings for mechanical systems}%
%

%

\author{Gen Najima}%
%

%

\author{Tsuneya Yoshida}%
%

%

\author{Yasuhiro Hatsugai}%
%

%

\affiliation{
Graduate School of Pure and Applied Sciences,
University of Tsukuba, Tsukuba, Ibaraki 305-8571, Japan\\
Department of Physics, University of Tuskuba,
Ibaraki 305-8571, Japan}%
\date{\today}%
%

\begin{abstract}%

Non-Hermitian topological phenomena occur in mechanical systems
described by the Newton equation.
A mechanical graphene,
which is composed of mass points and springs,
shows symmetry-protected exceptional rings (SPERs)
in the presence of the friction.
However, it remains unclear what
physical properties or phenomena the SPERs induce.
Our numerical analysis reveals that
the SPERs can govern dynamics in the wavenumber space.
Moreover,
we propose how to extract the dynamics in the wavenumber space
for systems with the boundaries.
Furthermore,
we also observe that
connectivity of the SPERs changes depending on the fiction,
which is analogous to the Lifshitz transition of electron systems.

%






%

\end{abstract}%
%

\maketitle

\section{Introduction}
Since topological insulators were realized
in HgTe/CdTe quantum wells
[\onlinecite{Kane2005,Ahn2006,Konig1980,PhysRevB.48.11851}],
topological phenomena have been
studied extensively [\onlinecite{PhysRevLett.71.3697,Hasan2010,Qi2011}].
Topological phenomena have been confirmed not only in quantum systems
but also in classical systems.
Typical examples are systems described by Maxwell's equations e.g.,
photonic crystals [\onlinecite{Haldane2008,Wang2009,Raghu2006}]
and electric circuits [\onlinecite{Ningyuan2015,Albert2015,Lee2018,PhysRevResearch.2.022062}].
Additionally, topological phenomena were reported for in
mechanical systems described by
Newton's equation [\onlinecite{S_sstrunk_2016,Kariyado2015a,Kane2013,Takahashi_2017,Takahashi2019,Wakao2020,Prodan_2009,Po_2016,Paulose2015,Rocklin_2016,Wang_2015,Socolar_2017}].
Specifically, it has been elucidated that
the topological phenomena
of quantum systems can be observed by
solving the equation of motion of
the mechanical graphene which is one of the spring-mass models.
Mathematically, the emergence of topological phenomena
beyond quantum systems can be understood from
the fact that the systems are described by
an eigenvalue problem of a Hermitian matrix.

More recently,
for non-Hermitian systems,
new topological phenomena have been reported
which do not have Hermitian counterparts
[\onlinecite{Bergholtz2019,Shen2018,Shen2018a,Ghatak2019,Lee2016,Kozii2017,Hatano1996,Bender1998,Fukui,Sone2019,Lee_2016,Gong2017,Guo2009,Ruter2010,Regensburger2012,Zhen2015,Hassan2017,Zhou,Kozii2017}].
For instance,
the non-Hermiticity induces exceptional points
(or its symmetry-protected variants)
where both of the real- and the imaginary-parts of
the energy bands touch
[\onlinecite{Carlstrom2018,Yoshida2018,Yoshida2019a,Budich2019,Okugawa2019,Zhou2019,Xu2017}].
A theoretical work [\onlinecite{Yoshida2019}] elucidated
the emergence of the symmetry-protected exceptional rings (SPERs) in
a mechanical graphene
with friction which is composed of mass points and springs.
However, the SPERs in mechanical systems
have not been experimentally observed yet.
One of the reasons for this is
that it is not clear how to observe the SPERs.
In addition,
theoretical discussion elucidating
the effects of boundaries is missing
although most of experiments are carried out for systems with boundaries.

In this paper,
we numerically analyze effect of SPERs on dynamics of
a mechanical graphene prior to the experimental studies.
Our analysis elucidates that SPERs can be observed by
examining the time-evolution of the displacements
of mass points in the wavenumber space.
We also elucidate
how to extract the data in the wavenumber space
for systems with boundaries.
Applying this approach, we find that
the SPERs can be observed by examining the time-evolution
before the oscillation reaches to the boundaries.
Furthermore, we also discover a phenomenon
which is analogous to
the Lifshitz transition in electron systems;
the connectivity of SPERs changes
depending on the friction coefficient.

The rest of this paper is organized as follows.
In Sec.~\ref{Sec:02}, we present an overview of
the mechanical graphene.
Additionally, we show the difference in the formulation
of the Fourier transform depending on
the presence or absence of the boundaries.
In Sec.~\ref{Sec:04}, we show that
the SPERs can govern dynamics in the wavenumber space.
In Sec.~\ref{Sec:05}, we show that phenomenon corresponding to
the Lifshitz transition in mechanical systems.

\section{Equation of motion of the mechanical graphene}
\label{Sec:02}
In this section,
we briefly review that
the time-evolution of the mechanical graphene is described by
the Newton equation which is mathematically equivalent to
the Schr\"odinger equation in the wavenumber space.
The obtained equation implies that the SPERs governs
the dynamical properties.

We also discuss how to extract the data
in the wavenumber space for systems with the boundaries,
motivated by the fact that
ordinary experiments are carried out for systems with the boundaries.

\begin{figure}[t]
  \begin{center}
    \includegraphics[scale=0.23]{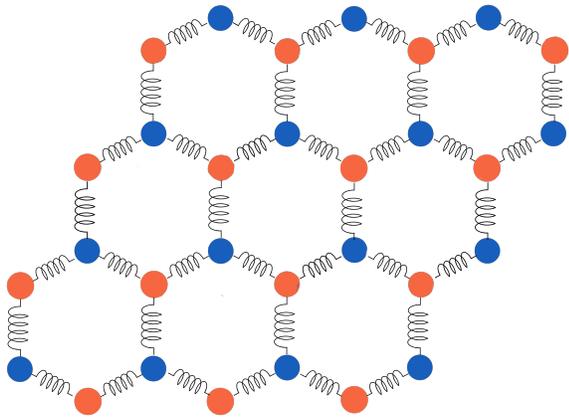}
    \caption{Schematic picture of the mechanical graphene.
    The system is composed of mass points and springs.
    The $A$ and $B$ sublattices are denoted by red and blue mass points,
    respectively.}
    \label{Fig:01}
  \end{center}
\end{figure}%

\subsection{Equation of motion without boundaries}

Let us consider a system under the periodic boundary condition.
The Fourier transformed displacement is written as
\begin{align}
  \label{eq:fourier}
  u_{\bm{k},\alpha}^{\mu} = \frac{1}{N}\sum_{\bm{R}}x_{\bm{R},\alpha}^{\mu}e^{i\bm{k}\cdot\bm{R}},
\end{align}
where $u_{\bm{k},\alpha}^{\mu}$ denotes the Fourier transformed displacement
along $\mu$-direction $(\mu,\nu=x,y)$.
$\alpha$ denotes sublattice $A, B$.
The number of unit cells are denoted by $N$.

By applying the Fourier transformation,
the equation of motion is written as,
\begin{align}
  \ddot{u}_{\bm{k},\alpha}^{\mu} = -D^{\mu\nu}_{\alpha\beta}(\bm{k})u^{\nu}_{\bm{k},\beta}
  - b\hat{1} \dot{u}^{\mu}_{\bm{k},\alpha},
\end{align}
with $\dot{u}^{\mu}_{\bm{k},\alpha} = \frac{d u^{\mu}_{\bm{k},\alpha}}{dt}$.
Summation over the repeated indices $\mu$, $\nu$ is assumed.
The first and second terms describe the potential force and
the frictional force proportional to the velocity respectively.
The matrix $D(\bm{k})$ includes the mass $m$ of the mass points,
the spring constant $\kappa$,
and the tension $\eta$ as parameters.
The explicit form of $D$ is given in Appendix \ref{App:01}.

In a matrix form, the above equation is written as
\begin{subequations}
\begin{align}
  \dot{\bm{\phi}}_{\bm{k}}(t) &= M(\bm{k})\bm{\phi}_{\bm{k}}(t),\\
  \label{eq:matrix}
  M(\bm{k}) &= \left(
  \begin{array}{cc}
    0 & \hat{1}\\
    -D(\bm{k}) & -b\hat{1}
  \end{array}
  \right),
\end{align}
\end{subequations}
with $\bm{\phi}_{\bm{k}} = (\bm{u}_{\bm{k}}, \dot{\bm{u}}_{\bm{k}})^T$.

We note that
the momentum region, where $M(\bm{k})$ cannot be diagonalized,
appears as a ring. This ring is knowns as an SPER
(for more details such as relevant symmetry of SPERs,
see Appendix \ref{App:sym}).
The above results imply that
the SPER may affect the dynamics,
which is explicitly demonstrated in Sec.~\ref{Sec:05}.

\subsection{dynamical properties for systems with boundaries}
In the previous section,
we have briefly reviewed the case
under the periodic boundary condition.
In this section, we describe how to obtain data
in the wavenumber space from that in the real space
when the system has the boundaries.

Let $x^\mu_{\bm{\tilde{R}}}$ be
the displacement of a mass point at position $\bm{\tilde{R}}$.
Then,
the time-evolution of $x^{\mu}_{\bm{\tilde{R}}}$ is obtained
from the equation of motion for $x^{\mu}_{\bm{\tilde{R}}}$
regardless of the presence/absence of the boundaries.
For systems with the boundaries,
the data in the wavenumber space is obtained by the Fourier integration.
By applying the Fourier integration to $x^\mu_{\bm{\tilde{R}}}$,
we can calculate the displacement
$u^{\mu}_{\bm{k}}$ in the wavenumber space,
\begin{align}
  \label{eq:B01}
  u^{\mu}_{\bm{k},\alpha} = \frac{1}{2\pi}\int_{-\infty}^{\infty}\int_{-\infty}^{\infty}d^2\bm{\tilde{R}}~x^{\mu}_{\bm{\tilde{R}},\alpha} e^{i\bm{k}\cdot\bm{\tilde{R}}},
\end{align}
Expressing $x^{\mu}_{\bm{\tilde{R}},\alpha}$
as a function of $\bm{\tilde{R}}$,
we can write it as a sum of delta functions,
\begin{align}
  x_{\bm{\tilde{R}},\alpha}^{\mu} = \sum_i x_{\bm{\tilde{R}}_i,\alpha}\delta(\bm{\tilde{R}} - \bm{R}_i).
\end{align}
Therefore, Eq.~(\ref{eq:B01}) can be expressed as,
\begin{align}
  \label{eq:fomu}
  u_{\bm{k},\alpha}^{\mu} = \sum_i x^{\mu}_{\bm{R}_i,\alpha}e^{i\bm{k}\cdot\bm{R}_i},
\end{align}
where the sum is taken for
all mass points belonging to sublattice $\alpha$.

Equation (\ref{eq:fomu}) is similar to Eq.~(\ref{eq:fourier}).
The difference between them is in the definition of wavenumber.
In the latter case, the wavenumber is discrete,
while in the former case, the wavenumber is continuous.

The difference between the free boundary and the fixed boundary
is determined when solving the equation of motion for
the displacement $x_{\bm{\tilde{R}}}$ in the real space.

\begin{figure}[tb]
  \begin{center}
    \includegraphics[scale=0.34]{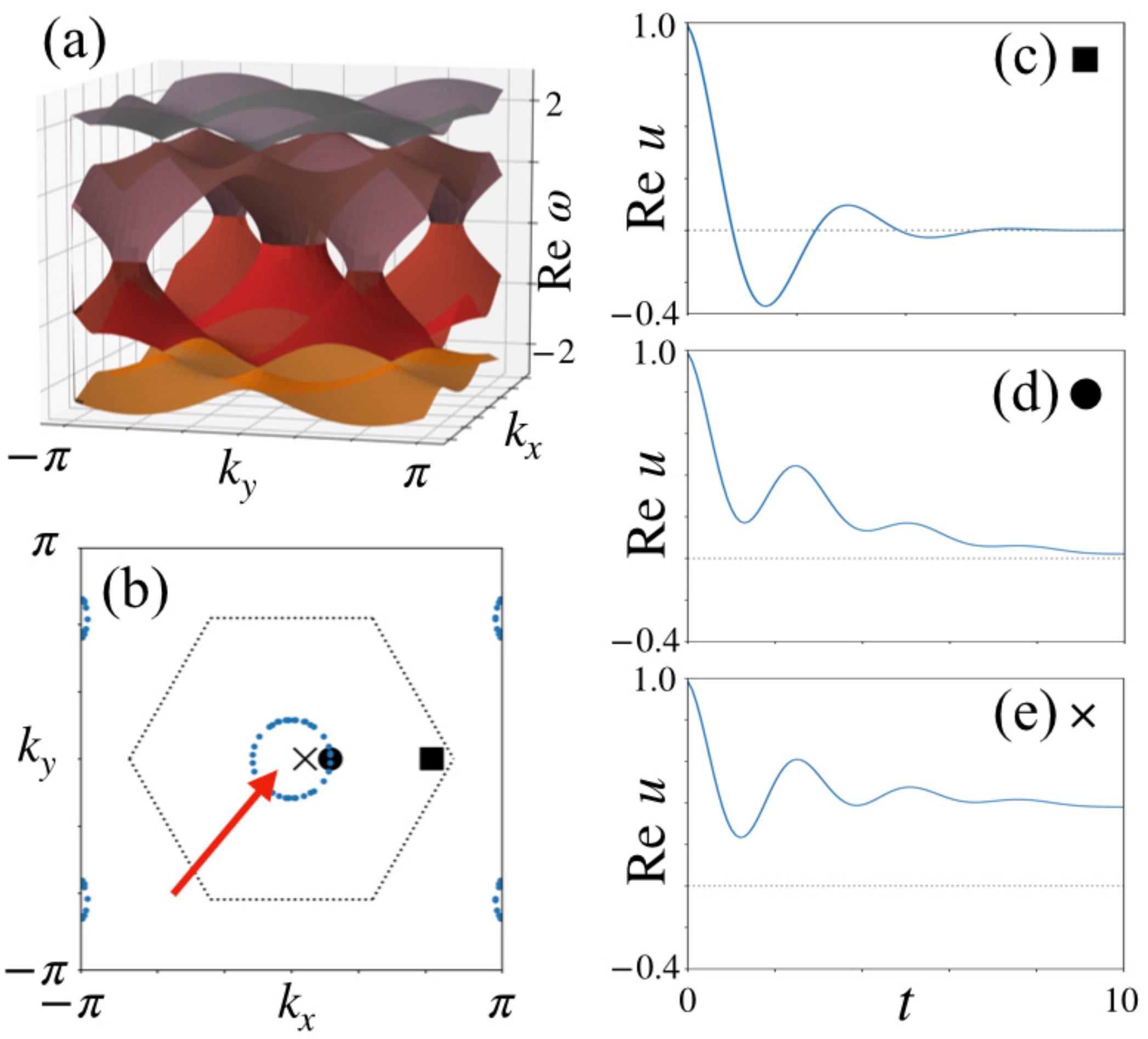}
    \caption{
    (a): The real-part of dispersion relation for the
    mechanical graphene.
    Band touching points correspond to the SPERs.
    (b): The SPERs for the mechanical graphene.
    Black dotted lines illustrate the Brillouin zone.
    (c)-(e): Time-evolution of $\mathrm{Re}u_{\bm{k}}$
    at the specific point in the Brillouin zone.
    (c): Outside of the SPERs, the system shows
    a underdamping.
    (d): On the SPERs, the system shows a critical damping.
    (e): Inside of the SPERs, a overdamping is observed;
    $\mathrm{Re}u_{\bm{k}}$ remains positive.
    The data shown in this figure
    are for $(m,\kappa,\eta,b)=(1, 1, 0, 1)$.
    }
    \label{Fig:03}
  \end{center}
\end{figure}%

\section{dynamics of systems without boundary}
\label{Sec:04}
In this section,
we elucidate that the SPERs can be observed
by examining the dynamics of displacements
in the wavenumber space.

\subsection{SPERs and types of damping}
By numerically diagonalizing the matrix $M(\bm{k})$,
we can obtain the dispersion relation $\omega(\bm{k})$ [see Fig.~2(a)].
In this figure, the band touching points form rings,
which corresponds to the SPERs.
The violation of the diagonalizability
on the SPERs can also be observed
in Fig.~\ref{Fig:03}(b).
This figure displays the momenta
where $M(\bm{k})$ cannot be diagonalized.

In Sec.~\ref{Sec:fb},
we provides the numerical data elucidating that
the SPERs govern the dynamics.
Here, prior to the numerical simulation,
we note the relation between $\omega(\bm{k})$ and
the dispersion relation
in the absence of the friction.

Applying a unitary transformation,
the matrix $M(\bm{k})$ is rewritten as
\begin{align}
  \label{Eq:diag}
  \tilde{U}^\dagger(\bm{k}) M(\bm{k}) \tilde{U}(\bm{k}) =
  \bigoplus_n
  \left(
    \begin{array}{cc}
      0 & 1 \\
      -\omega_{0n}^2 & -b
    \end{array}
  \right),
\end{align}
where $\tilde{U}(\bm{k})$ is an unitary matrix (see Appendix \ref{App:02}).
Because the matrix
appearing in the left-hand side of Eq.~(\ref{Eq:diag})
is a set of $2\times2$-matrix,
we obtain
\begin{align}
  \label{omegan}
  \omega_n(\bm{k}) = -\frac{b}{2}i \pm \sqrt{\omega^2_{0n}(\bm{k}) - \left(\frac{b}{2}\right)^2},
\end{align}
where $\omega^2_{0n}$ ($n=1,2,\ldots,\mathrm{dim}D$) denote
eigenvalues of the matrix $D(\bm{k})$.
Equation~(\ref{omegan}) indicates
that the eigenvalues $\omega(\bm{k})$ take
complex values inside of the SPERs.
On the other hand, outside of the SPERs, $\omega(\bm{k})$ take
real values.
It means that
the Fourier transformed displacement $u_{\bm{k}}^{\mu}$
behaves differently inside and outside of the SPERs.
Specifically, the underdamping (overdamping) is observed outside
(inside) of the SPERs.
Additionally, on the SPERs,
$u_{\bm{k}}$ become the critical damping,
which converges most quickly.
In Sec.~\ref{Sec:fb}, we discuss the above behaviors in details.

\subsection{Numerical demonstration of dynamics}
We demonstrate that
the SPERs separates regions of overdamping
and regions of underdamping.
This fact indicates that the SPERs govern the dynamics.

In the following,
we set $u_{\bm{k}}(t=0)$ to $u_{\bm{k}}(t=0)=1$ as the initial condition.
This condition corresponds to the case
where the mass point at the origin is displaced at $t=0$.

\begin{figure}[tb]
  \begin{center}
    \includegraphics[scale=0.4]{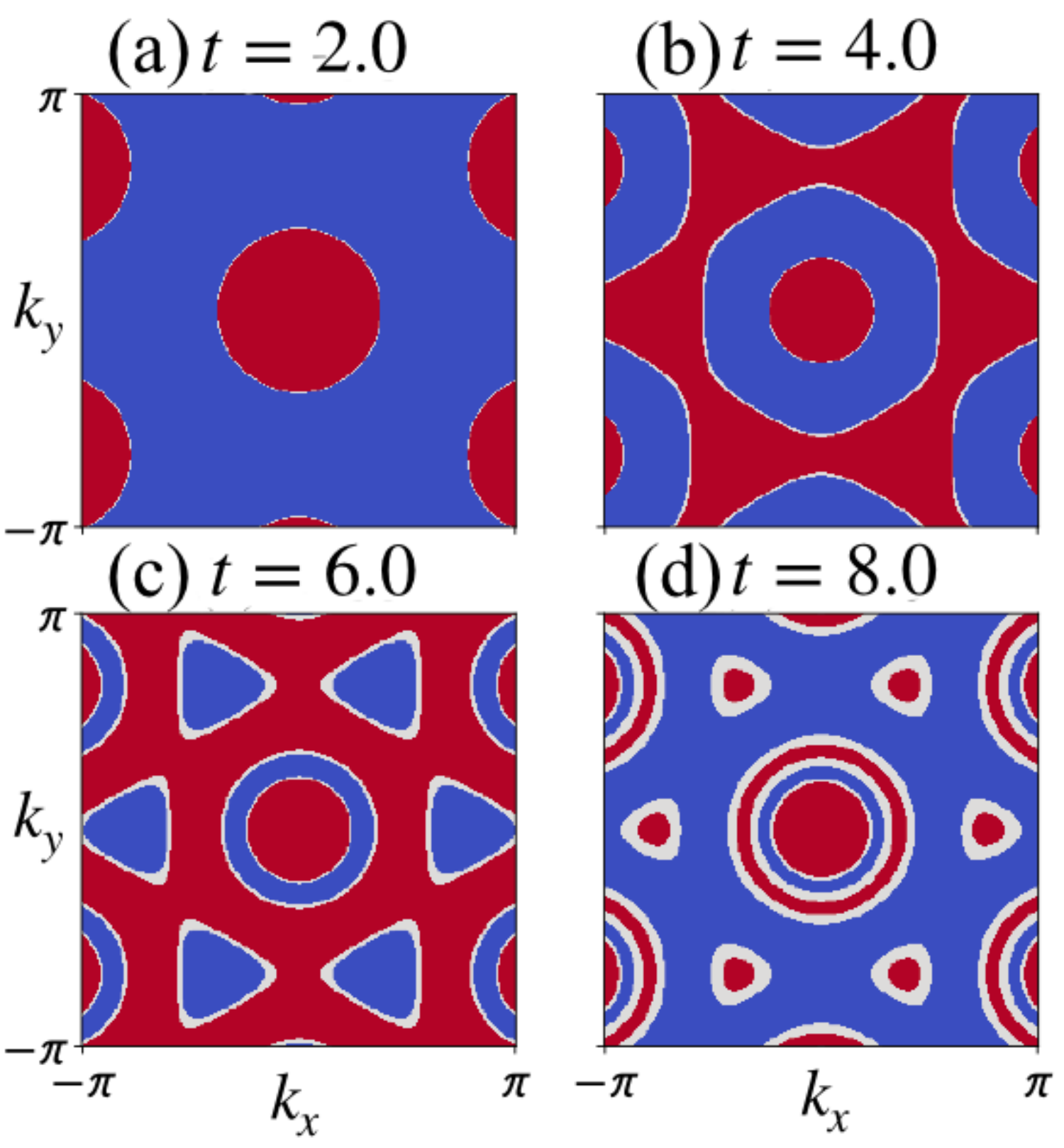}
    \caption{Snapshots of time-evolution
    of $\mathrm{Re}u_{\bm{k}}$ over all wavenumber
    (initial condition is $u_{\bm{k}}$ = 1 for all $\bm{k}$).
    In red (blue) regions, sign of $\mathrm{Re}u_{\bm{k}}$ is plus (minus).
    White region means $|\mathrm{Re}u_{\bm{k}}| < 0.005$.
    Inside the SPERs, the sign remains positive.
    In the region around the $\Gamma$ point
    [i.e., $\bm{k}=(0,0)$], the sign of displacement is unchanged
    up to $t=6.0$.
    The boundary of this region corresponds to the SPER.}
    \label{Fig:04}
  \end{center}
\end{figure}%

\subsubsection{Specific wavenumber}

Let us first discuss the time-evolution for given momenta.
Figures \ref{Fig:03}(c)-\ref{Fig:03}(e) indicate
the dynamics of displacement for given wave numbers
denoted by a cross,
a cycle, and a square in Fig.~\ref{Fig:03}(b).

The contribution of the time-evolution of $u_{\bm{k}}$
can be expressed as
$e^{i\omega(\bm{k}) t} = e^{-\mathrm{Im}\omega(\bm{k})\cdot t}\cdot e^{i\mathrm{Re}\omega(\bm{k})\cdot t}$.
Thus, in addition to the oscillation,
damping occurs
as $\omega$ takes a complex value [see Fig.~\ref{Fig:03}(c)].
As we approach closer to the inner side of
the SPER [i.e., the region denoted by
the arrow in Fig.~\ref{Fig:03}(b)],
the oscillations reach a critical damping
on the SPER, where $\omega$ becomes purely imaginary
[see Fig.~\ref{Fig:03}(d)].
Inside of the SPER,
overdamping is observed
[see Fig.~\ref{Fig:03}(e)].
Here, we note that oscillation is observed even
inside of the SPERs and on the SPERs.
This is because the gray and orange bands in
Fig.~\ref{Fig:03}(a) are also involved in the dynamics.

The displacement insides of the SPERs decays slower
than that on the SPERs.
This behavior can be understood as follows.
Inside of the SPERs, we have two eigenvalues
$\omega_{n}= -ib/2 \pm i\sqrt{|\omega^2_{0n}(\bm{k}) -\frac{b}{2}|}$.
In contrast, on the SPERs, we have $\omega=-ib/2$ due to band touching.
Because the eigenmodes with
$\omega_{n}= -ib/2 + i\sqrt{|\omega^2_{0n}(\bm{k}) -\frac{b}{2}|}$
decays slower than the modes with $\omega_n=-ib/2$,
the displacement inside of the SPERs decays slower than that on the SPERs.

\subsubsection{Over all wavenumber}

The time-evolution of
the displacement in the wavenumber space is shown in Fig.~\ref{Fig:04}.
This figure indicates that $u_{\bm{k}}$ changes its sign.
The red (blue) region means $u_{\bm{k}}$ takes
positive (negative) values.
The white region means $u_{\bm{k}} = 0$.
We confirm that inside of the SPERs $u_{\bm{k}}$
remains positive even if time passes.
while outside of the SPERs,
the sign of $u_{\bm{k}}$ changes as time passes.

This result indicates that
the SPERs for the mechanical graphene
can be observed
by examining the
time-evolution in the wavenumber space.

\begin{figure}[tb]
  \begin{center}
    \includegraphics[scale=0.31]{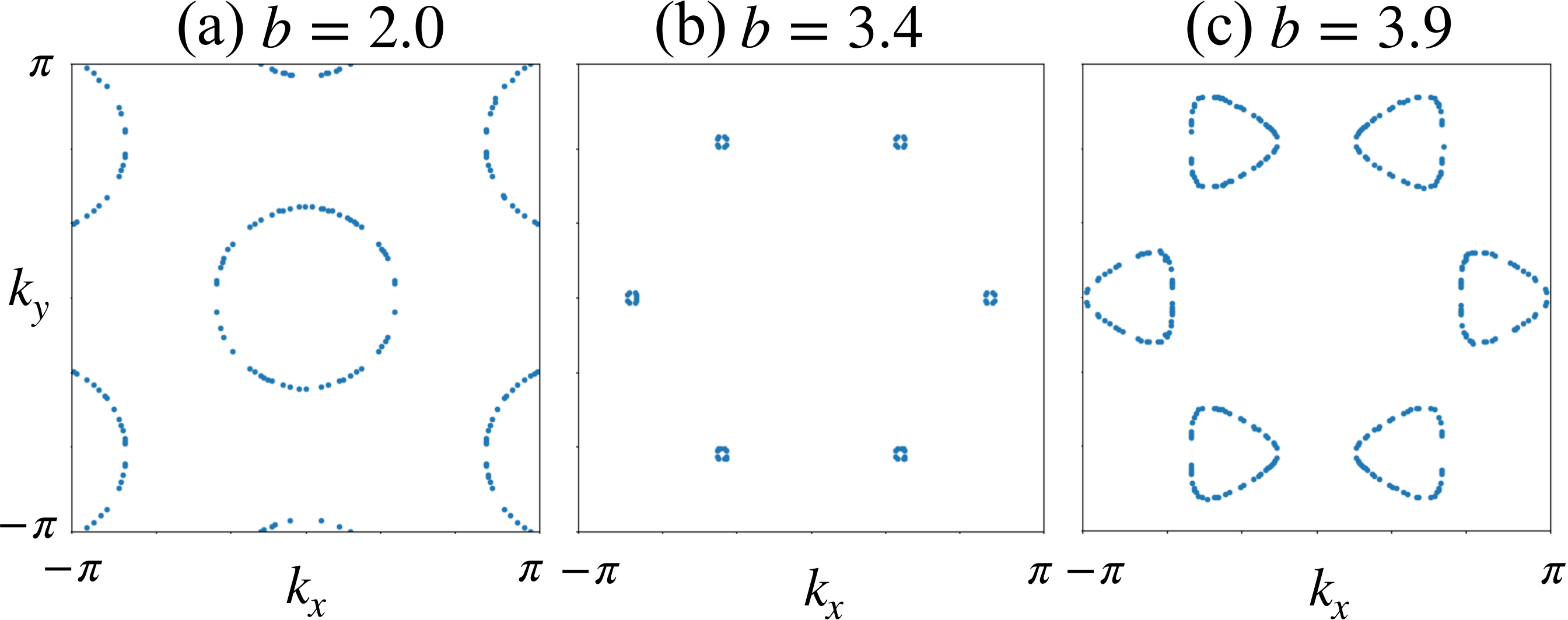}
    \caption{
    (a), (b) and (c): The SPERs for $b=2.0$, $3.4$ and $ 3.9$,
    respectively. The data shown in this figure
    are obtained for momentum points $\bm{k}_0$ satisfying
    $|\mathrm{det}M(\bm{k}_0)| < 0.005.$
    }
    \label{Fig:Lifshitz_transition}
  \end{center}
\end{figure}%
\section{Lifshitz transition for mechanical systems}
\label{Sec:05}

In this section,
we point out that the mechanical graphene with the friction
exhibits a behavior
which is analogous to the Lifshitz transition in quantum systems.
The Lifshitz transition is known as
changes of the connectivity of
the Fermi surface for electronic systems at zero temperature;
shifting the Fermi energy changes the connectivity of
the Fermi surface.
Here, for given Fermi energy, the Fermi surface appears
as a cross section of the bands and the Fermi energy.

\subsection{Periodic boundary condition}
Firstly, we point out that
the SPERs in the mechanical system are similar to
the Fermi surface of an electron system.
This correspondence can be seen in Eq.~(\ref{omegan}).
This equation indicates that the SPERs emerge as cross section
of the dispersion relation $\omega_{0n}(\bm{k})$
and half of the frication coefficient $b/2$.
Specifically, at momenta satisfying,
\begin{align}
  \label{Eq:omega-b}
  \omega_{0n}(\bm{k}) = \frac{b}{2},
\end{align}
$M(\bm{k})$ becomes exceptional.

Therefore, with increasing the frictional coefficient,
the connectivity of the SPERs changes.
Figure \ref{Fig:Lifshitz_transition} plots the SPERs for several values
of the frictional coefficient $b$.
This figure indicates that with increasing $b$ from $b=2.0$,
the SPERs shrinks to points
[see Fig.~\ref{Fig:Lifshitz_transition}(a)
and \ref{Fig:Lifshitz_transition}(b)].
Further increasing $b$, the SPERs  emerges again
[see Fig.~\ref{Fig:Lifshitz_transition}(c)].
This change of the connectivity is analogous to
the Lifshitz transition of electron systems.

\begin{figure}[tb]
  \begin{center}
    \includegraphics[scale=0.35]{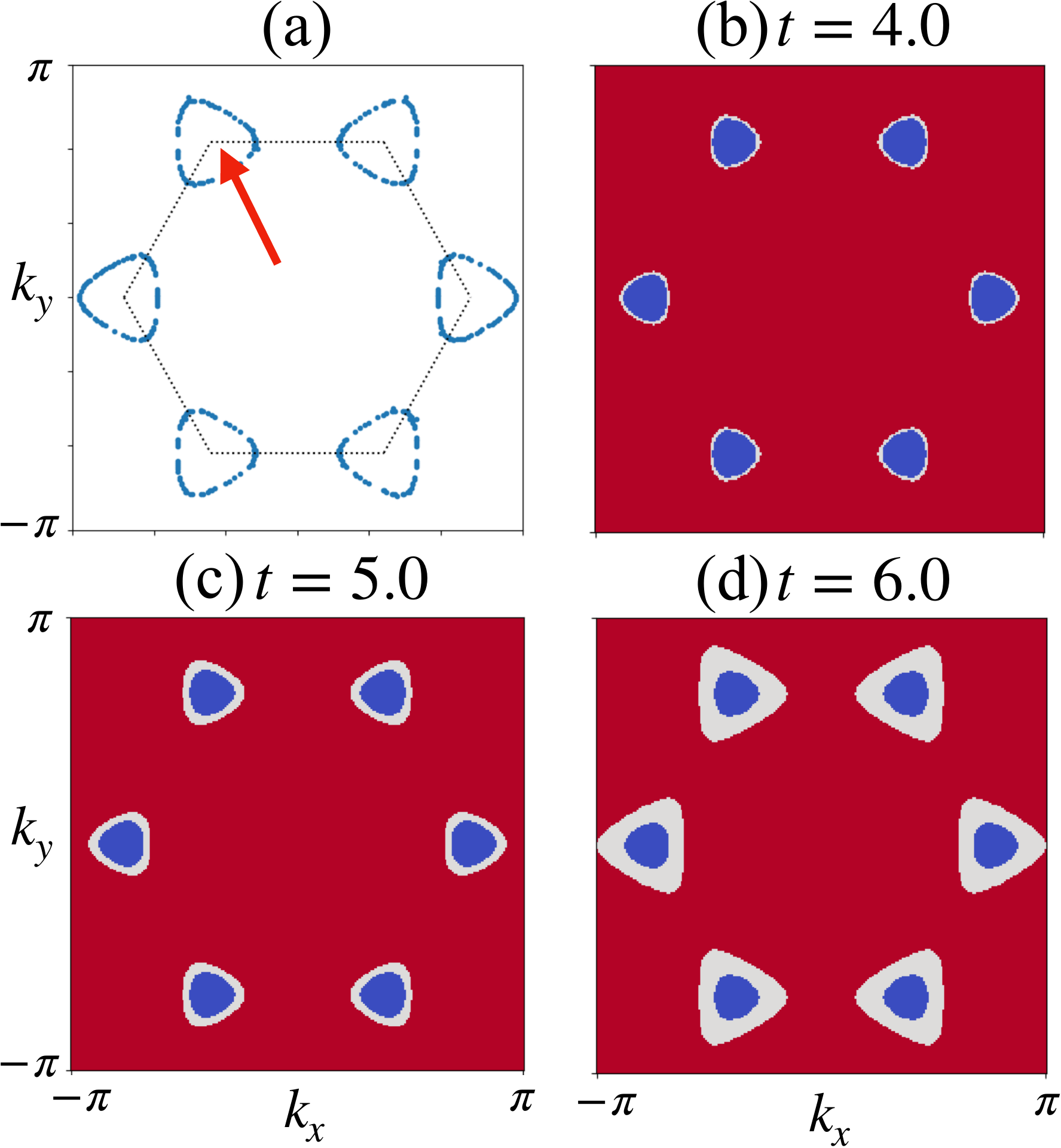}
    \caption{Dynamics of the Fourier transformed
    displacement $u_{\bm{k}}$ for $b=3.0$ with the periodic boundary.
    (a): The SPERs for $b=3.0$.
    (b), (c), and (d): Snapshots of time-evolution
    of $u_{\bm{k}}$ over all wavenumber
    (initial condition is $u_{\bm{k}}$ = 1 for all $\bm{k}$)
    at $t=4.0$, $5.0$, and $6.0$, respectively.
    In the region around the $K$ point
    [i.e., $\bm{k}=(2\pi/3\sqrt{3},2\pi/3)$],
    the sign of the displacement is unchanged
    up to $t=6.0$.
    The boundary of this region corresponds to the SPER.}
    \label{Fig:Lifshits_period}
  \end{center}
\end{figure}

The above ``Lifshitz transition'' in the mechanical system
also affects the dynamical properties.
Figure \ref{Fig:Lifshits_period} shows that dynamics of $u_{\bm{k}}$
for $b=3.0$ under the periodic boundary condition.
This figure shows that
$u_{\bm{k}}$ behaves the overdamping
(underdamping) outside (inside) of the SPERs.
This behavior for $b=3.0$ is in sharp contrast
to that for $b=1.0$.
For $b=3.0$, the underdamping is observed
inside of the SPERs
[i.e., the region denoted by
the arrow in Fig.~\ref{Fig:Lifshits_period}(a)]
while the overdamping is observed
outside of the SPERs.
Namely, compared to the case for $b=1.0$,
the region of overdamping and that of the
underdamping are exchanged.

In the above,
we have seen that connectivity of the SPERs changes
depending on the frictional force,
which is analogous to the Lifshitz transition of electron systems.
Our numerical simulation has elucidated that
the change of the connectivity also affects the dynamical properties.

\subsection{Fixed boundary condition}
\label{Sec:fb}

So far, we have analyzed
the system under the periodic boundary condition.
In this section,
we investigate the
``Lifshitz transition"
for systems with the boundaries
because most of experiments
are carried out for systems with boundaries.

\begin{figure}[tb]
  \begin{center}
    \includegraphics[scale=0.3]{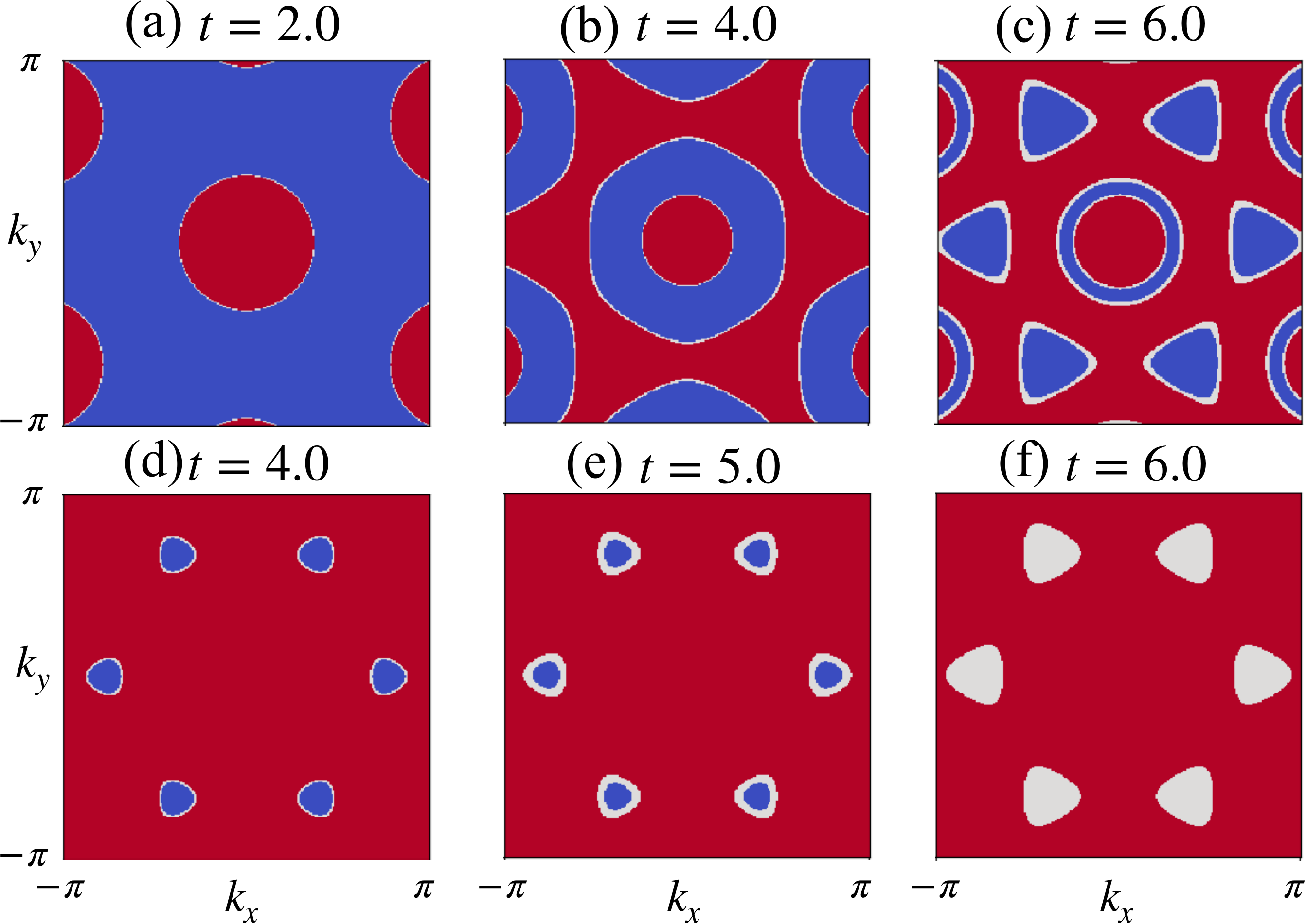}
    \caption{(a), (b), and (c): [(d), (e), and (f):]
    Snapshot of the dynamics of $u_{\bm{k}}$
    at $t=2.0$, $4.0$, and $6.0$
    [$t=4.0$, $5.0$, and $6.0$] with
    the fixed boundary condition
    at $b=1.0$ [$b=3.0$] respectively.
    One can see that changing the value of $b$ causes
    the dynamics to switch
    between the inside and outside of the SPERs
    (initial condition is $u_{\bm{k}}$ = 1 for all $\bm{k}$).
    In panels (a)-(c), the sign of displacement is unchanged
    up to $t=6.0$ for the region around the $\Gamma$ point
    [i.e., $\bm{k}=(0,0)$].
    The boundary of this region corresponds to the SPER.
    In panels (d)-(f),
    the sign of the displacement is unchanged
    up to $t=6.0$
    for the region around the $K$ point
    [i.e., $\bm{k}=(2\pi/3\sqrt{3},2\pi/3)$].
    The boundary of this region corresponds to the SPER.}
    \label{Fig:Lifshits_boundary}
  \end{center}
\end{figure}%

Figure \ref{Fig:Lifshits_boundary} shows
the results of the numerical simulation
of the mechanical graphene
with the fixed boundary condition.
Specifically,
the numerical data are obtained for the system composed of 242 sites
(for more details see
Fig.~\ref{Fig:rhombus} of Appendix~\ref{App:03}).
For $b=1.0$,
the sign of $u_{\bm{k}}$ does not change
during the time-evolution in the region surrounded by the SPERs
[see Figs. \ref{Fig:Lifshits_boundary}(a)-\ref{Fig:Lifshits_boundary}(c)].
In other words, the overdamping occur in this region.
For $b=3.0$,
it can be seen that the region of the overdamping expands.
The above results are consistent with the results
without the boundaries, which indicates the ``Lifshitz transition"
is also observed for the system under the fixed boundary conditions
by examining the dynamics of the displacement in the
wavenumber space.

The results with and without the boundaries are consistent
when the oscillations do not reach to the boundaries of the system.
When the oscillations propagate to the boundaries,
the displacement in the wavenumber space
behaves differently from
the results for the periodic boundary condition
due to the influence of reflected waves.
Therefore, in order to observe the SPERs,
the size of the system must be large enough
so that the effect of the boundaries can be neglected
(see Appendix~\ref{App:06}).
Although Fig.~\ref{Fig:Lifshits_boundary} shows
the results for the fixed boundary condition,
the same results can be obtained for the free boundary condition,
since the results are not affected by the reflected wave
unless the oscillation reaches to the boundaries.

\section{Summary}
Despite the previous theoretical prediction of the SPERs
in mechanical systems,
the experimental observation has not been accomplished yet.
In this paper, prior to experimental realization,
we have theoretically addressed the following issues:
(i) how the emergence of the SPERs affects dynamical behaviors
which are experimentally accessible;
(ii) what is the proper formulation of the Fourier transform
in the systems with the boundaries.

Our results elucidate that the SPERs for the mechanical graphene
govern the dynamics in the wavenumber space.
Outside of the SPERs,
the Fourier transformed displacement $u_{\bm{k}}$
shows the underdamping, while inside of the SPERs,
$u_{\bm{k}}$ shows the
overdamping (i.e., the sign of $u_{\bm{k}}$ remains plus).
On the SPERs, critical damping is observed,
which switches from the overdamping to the underdamping.

Furthermore, we have observed the phenomenon
corresponding the Lifshitz transition of electron systems.
Increasing the friction coefficient,
connectivity of the SPERs changes.
This change of the connectivity,
analogous to the Lifshitz transition, is observed
by examining the dynamical properties.

\section{Acknowledgements}
This work is partly supported
by JSPS KAKENHI Grants
No.~17H06138, No.~JP20H04627, and No.~JP21K13850.

\bibliographystyle{apsrev4-1}
%

\newpage

\appendix

\section{Details of the mechanical graphene}
\label{App:01}
\begin{figure}[t]
  \begin{center}
    \includegraphics[scale=0.3]{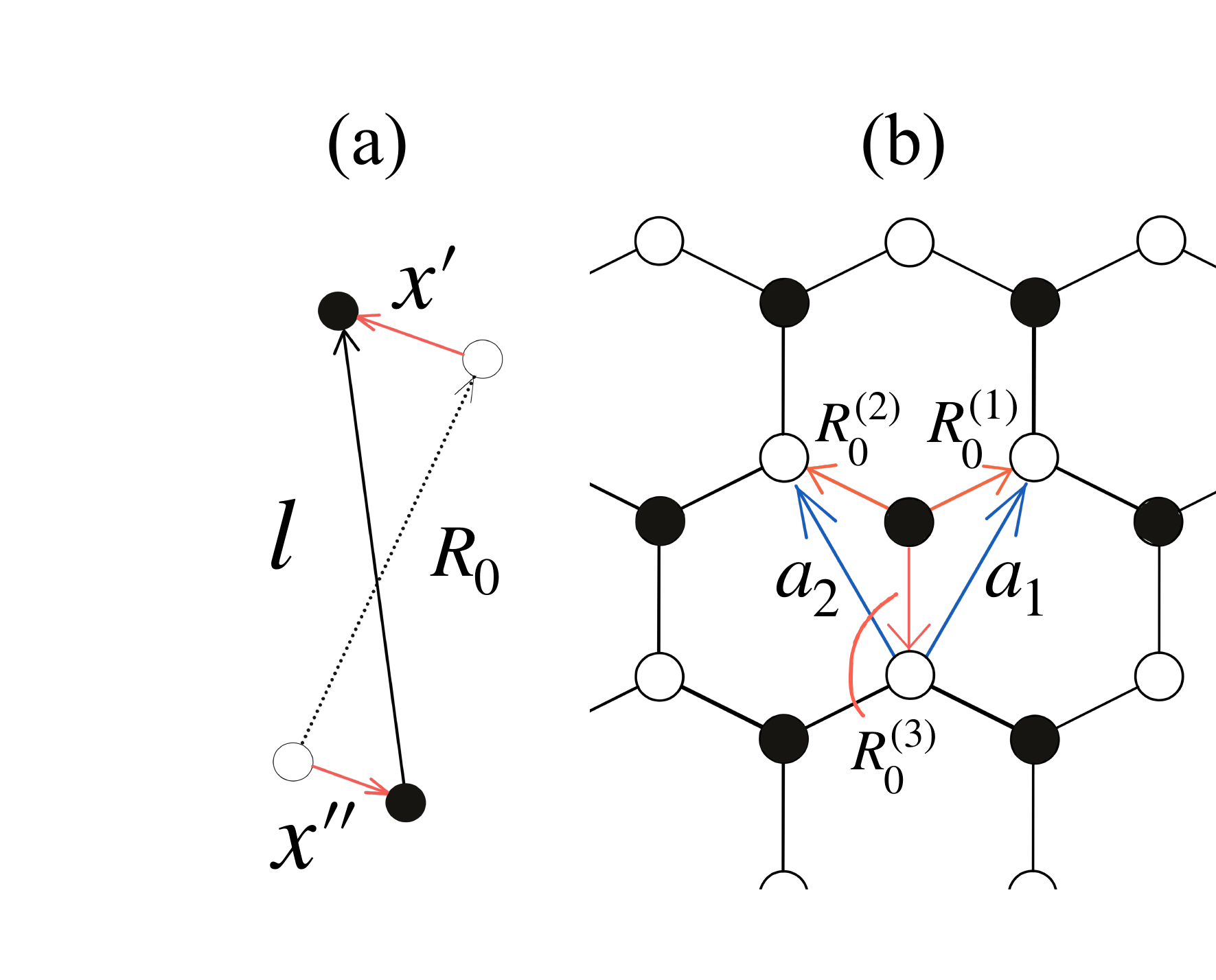}
    \caption{(a): Definitions of $\bm{R}_0$, $\bm{x'}$, and $\bm{x''}$.
    (b): Definitions of unit vectors $\bm{a}_1$ and $\bm{a}_2$, the vectors
    connecting the nearest neighbor mass points $\bm{R}_0$. }
    \label{Fig:unit}
  \end{center}
\end{figure}%

The model treated in this paper is the mechanical graphene,
which is composed of mass points and springs (see Fig.~\ref{Fig:01}).
Mass points are aligned in honeycomb lattice.
Here, we obtain the motion of equation of the mechanical graphene.

\subsection{In case of no friction}
First, to find the Lagrangian of the system,
we focus the unit cell which is plotted in Fig.~\ref{Fig:unit}.
Then, the elastic energy is written as
\begin{align}
  U_s = \frac{1}{2}\kappa (l - l_0),
\end{align}
where $l$ is the length of the spring at the moment, and $l_0$ is
the natural length of the spring.
If $U_s$ is expanded to the second order
in $\delta\bm{x} = \bm{x}' - \bm{x}''$,
we get
\begin{align}
  U_s &= \frac{1}{2}\kappa\left(
    (R_0 - l_0)^2 + \right. \nonumber \\
    & \quad \left. 2(R_0-l_0)\hat{\bm{R}}_0\cdot\delta\bm{x} +
    \delta x_{\mu}\gamma_{\hat{\bm{R}}_0}^{\mu\nu}\delta x_{\nu}
  \right),\\
  \gamma^{\mu\nu}_{\hat{\bm{R}}_0} &= (1 - \eta)\delta^{\mu\nu} +
  \eta\hat{R}_0^{\mu}\hat{R}_0^{\nu},
\end{align}%
with $\bm{\hat{R}}_0 = \bm{R}_0 /|\bm{R}_0| $.
Summation over the repeated indices $\mu$, $\nu$ is assumed.

Then, the Lagrangian of this system becomes
\begin{align}
  L &= T - V,\\
  T &=
   \frac{1}{2}\sum_{\bm{R}a}\dot{x}^{\mu}_{\bm{R}a}\dot{x}^{\mu}_{\bm{R}a},\\
  V &= \frac{1}{2}\kappa\sum_{\langle\bm{R}'a\bm{R}b\rangle}(x^{\mu}_{\bm{\bar{R'}}a}
  - x^{\mu}_{\bm{\bar{R}}b} )\gamma^{\mu\nu}_{\bm{\bar{R'}}a - \bm{\bar{R}}b}
  (x^{\nu}_{\bm{\bar{R'}}a}
  - x^{\nu}_{\bm{\bar{R}}b} ),
\end{align}
where $\bm{\bar{R}_a}$ is a position of the sublattice associated with
the lattice point at $\bm{R}$.
Summation over $\langle\bm{R}'a\bm{R}b\rangle$ means to
take sum with respect to adjacent qualities.
Assuming the periodic boundary condition,
after the Fourier transformation,
\begin{align}
  x_{\bm{R}a} = \frac{1}{N}\sum_{\bm{k}}e^{i\bm{k}\cdot\bm{R}}u_{\bm{k}a}^{\mu},
\end{align}
the Lagrangian is rewritten as
\begin{align}
  L &= \frac{1}{N}\sum_{\bm{k}}L_{\bm{k}},\\
  L_{\bm{k}} &= \frac{1}{2}\sum_a \dot{u}^{\mu}_{\bm{k}a}\dot{u}^{\mu}_{-\bm{k}a}
  - \frac{1}{2}\sum_{ab}D^{\mu\nu}_{ab}u_{\bm{k}a}u_{-\bm{k}b},
\end{align}
where $D^{\mu\nu}_{ab} = D(\bm{k})$,
\begin{subequations}
\begin{align}
  D(\bm{k}) &= 3\kappa\left(
    1 - \frac{\eta}{2}
  \right) \hat{1}
  + \left(
    \begin{array}{cc}
      0 & D_{AB}(\bm{k})\\
      D_{AB}^{\dagger}(\bm{k}) & 0
    \end{array}
  \right),\\
  D_{AB}(\bm{k}) &= \kappa(\gamma_1 e^{-i\bm{k}\cdot\bm{a}_1} +
  \gamma_2 e^{-i\bm{k}\cdot\bm{a}_2} + \gamma_3),\\
  \gamma_1 &= (1-\eta)\left(
    \begin{array}{cc}
      1 & 0\\
      0 & 1
    \end{array}
  \right) + \eta\left(
    \begin{array}{cc}
      \frac{3}{4} & \frac{\sqrt{3}}{4}\\
      \frac{\sqrt{3}}{4} & \frac{3}{4}
    \end{array}
  \right),\\
  \gamma_2 &= (1-\eta)\left(
    \begin{array}{cc}
      1 & 0\\
      0 & 1
    \end{array}
  \right) + \eta\left(
    \begin{array}{cc}
      \frac{3}{4} & -\frac{\sqrt{3}}{4}\\
      -\frac{\sqrt{3}}{4} & \frac{3}{4}
    \end{array}
  \right),\\
  \gamma_3 &= (1-\eta)\left(
    \begin{array}{cc}
      1 & 0\\
      0 & 1
    \end{array}
  \right) + \eta\left(
    \begin{array}{cc}
      0 & 0\\
      0 & 1
    \end{array}
  \right),
\end{align}
\end{subequations}

Then, the equation of motion is given by
\begin{align}
  \ddot{u}_{\bm{k}}^{\mu} = -D^{\mu\nu}_{\alpha\beta}(\bm{k})u^{\nu}_{\bm{k}}.
\end{align}

\subsection{In case of friction}
Next, we consider the case of the friction.
Then, the equation of motion is rewritten as
\begin{align}
  \ddot{u}_{\bm{k}}^{\mu} = -D^{\mu\nu}_{\alpha\beta}(\bm{k})u^{\nu}_{\bm{k}}
  + \Gamma_{0\alpha\beta}^{\mu\nu}(\bm{k})\dot{u}^{\nu}_{\bm{k}},
\end{align}
The first and second terms describe the potential force and the
frictional force proportional to the velocity respectively.

In the matrix form, the above equation is rewritten as
\begin{subequations}
\begin{align}
  \label{eq:A13}
  \dot{\bm{\phi}}_{\bm{k}}(t) &= M(\bm{k})\bm{\phi}_{\bm{k}}(t),\\
  M(\bm{k}) &= \left(
  \begin{array}{cc}
    0 & \hat{1}\\
    -D(\bm{k}) & \Gamma_0(\bm{k})
  \end{array}
  \right),
\end{align}
\end{subequations}
with $\bm{\phi}_{\bm{k}} = (\bm{u}_{\bm{k}}, \dot{\bm{u}}_{\bm{k}})^T$.

When the friction is homogeneous,
matrix $\Gamma_0(\bm{k})$ is represented by $\Gamma_0(\bm{k}) = -b\hat{1}$.

In this way, the mechanical graphene with
the friction can be represented by the
eigenvalue equation of non-Hermitian matrix.
Practically, we can get eigenvalues by diagonalizing $M(\bm{k})$.

\subsection{Numerical demonstration}
\begin{figure}[t]
  \begin{center}
    \includegraphics[scale=0.4]{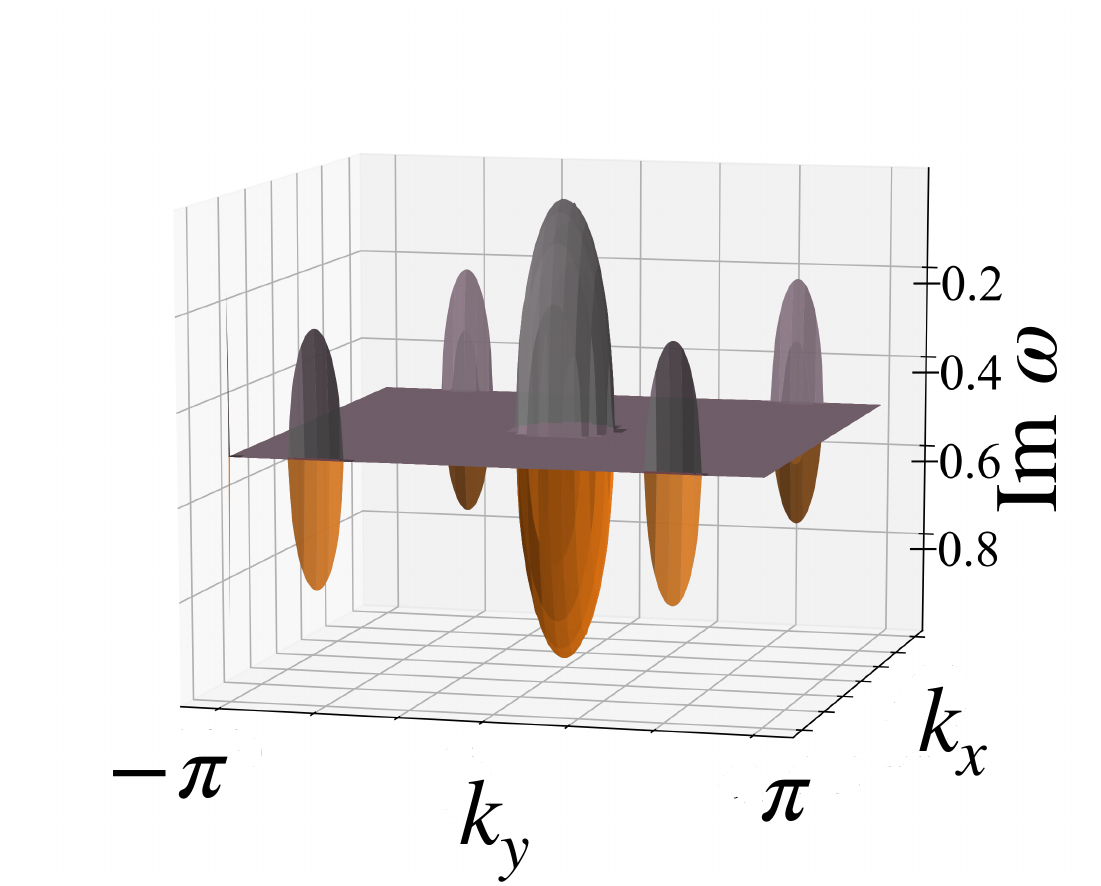}
    \caption{The imaginary-part of dispersion relation for the
    mechanical graphene.}
    \label{Fig:02}
  \end{center}
\end{figure}%

We demonstrate that the mechanical graphene hosts the SPERs.
By diagonalizing the Matrix $M(\bm{k})$, we can obtain
the dispersion relations $\omega(\bm{k})$ for $\eta=0$ and $b=1$
which is plotted in Fig.~\ref{Fig:03}(a) and Fig.~\ref{Fig:02}.
We can see that
rings of band touching points, which is caused by the friction.
At the rings of band-touching points, the Hamiltonian becomes defective.
In other words, the SPERs emerge for the mechanical graphene.
The emergence of SPERs can also be seen in Fig.~\ref{Fig:03}(b).
This figure shows that wavenumber points in the Brillouin zone
where the Hamiltonian is defective.
In other words, these points mean that
the determinant of the matrix $U(\bm{k})$ becomes zero.
Here, $U(\bm{k})$ is the matrix that lists
all eigenvectors of $M(\bm{k})$.
The SPERs are characterized by the topological quantity called
0th Chern number.

\section{Relation between symmetry and EPs}
\label{App:sym}
When there is symmetry in the system, the EPs change the form.
In this section, we discuss the symmetry of the mechanical system.
Additionally, we also show that EPs become the rings by symmetry.

\subsection{Hamiltonian of the mechanical graphene}
In order to discuss the symmetry,
we describe the corresponding Hamiltonian in
the mechanical graphene.

Because the matrix $D(\bm{k})$ is Hermitian and positive semidefinite,
using the Hermitian matrix $Q(\bm{k})$, it can be written as follows
\begin{align}
    D(\bm{k}) = Q^2(\bm{k}).
\end{align}
Therefore, Eq.~(\ref{eq:A13}) can be rewritten as
\begin{subequations}
\begin{align}
  \label{eq:shrodinger}
  i\frac{\partial}{\partial t}\psi(t) &= H(\bm{k})\psi(t),\\
  \label{eq:hamiltonian}
  H(\bm{k}) &= \left(
    \begin{array}{cc}
      0 & Q(\bm{k})\\
      Q(\bm{k}) & i\Gamma_0(\bm{k})
    \end{array}\right),
\end{align}
with
\begin{align}
  \psi(t) &= \left(
    \begin{array}{cc}
      Q(\bm{k}) & 0\\
      0 & i\hat{1}
    \end{array}
  \right)\phi(t).
\end{align}
\end{subequations}
The matrix $H(\bm{k})$ and $M(\bm{k})$ in Eq.~(\ref{eq:A13}) have
the same eigenvalues.
Therefore, the dynamics of the system can be described
by the matrix $H(\bm{k})$.
In addition,
since Eq.~(\ref{eq:shrodinger}) is essentially the same
as the Schr\"{o}dinger equation,
the matrix $H(\bm{k})$ can be regarded as Hamiltonian of the system.
In the following, we use Hamiltonian $H(\bm{k})$ to discuss symmetry.

When $i\Gamma(\bm{k})$ is Hermitian
(e.g. $\Gamma_0(\bm{k}) = -b\hat{1}$),
the system has extended chiral symmetry,
\begin{align}
  \label{eq:ecs}
  \tau_3 H^{\dagger}(\bm{k}) \tau_3 = -H(\bm{k}),
\end{align}
regardless of its detailed form
($\tau$'s are Pauli matrices).

\subsection{Symmetry and exceptional points}
Here we show that
in a two-dimensional system,
EPs form rings when there is a symmetry such as Eq.~(\ref{eq:ecs}).

For simplicity,
we consider the case
where the system can be described by $2\times 2$ Hamiltonian $H(\bm{k})$.
In general, $2\times 2$ matrix can be written
using Pauli matrices $\tau$'s as　
\begin{align}
  H(\bm{k}) = \sum_{\mu = 0,1,2,3}\left[ b_{\mu}(\bm{k}) + id_{\mu}(\bm{k}) \right]\tau_{\mu},
\end{align}
with $b_{\mu}, d_{\mu} \in \mathbb{R}~(\mu = 0,1,2,3)$.
The eigenvalues of $H(\bm{k})$ are as follows
\begin{align}
  \label{eq:eigen_sym}
  E_{\pm} = b_0 + id_0 \pm \sqrt{b^2 - d^2 + 2i\bm{b}\cdots\bm{d}},
\end{align}
with $\bm{b} = (b_1, b_2, b_3)^{T}$
and $\bm{d} = (d_1, d_2, d_3)^{T}$.
As can be seen from Eq.~(\ref{eq:eigen_sym}),
the band of the system touches the point
where the content of the root number becomes zero.
That is, the band touches at the points
which the following two equations hold
\begin{subequations}
\begin{align}
  b^2 - d^2 = 0,\\
  \label{eq:condition}
  \bm{b}\cdot\bm{d} = 0.
\end{align}
\end{subequations}
Since the system has two dimensions
and there are two conditionals,
$\bm{k}$ satisfying the conditionals
appears as zero-dimensional objects (points).
At this points, if we do not apply fine tune,
the Hamiltonian $H(\bm{k})$ is no longer diagonalizable.
That is, these points are EPs.

Next, we consider the case
where the system has symmetry as Eq~(\ref{eq:ecs}).
In this case, the Hamiltonian can be expressed as
\begin{align}
  H(\bm{k}) = id_0\tau_0 + b_1\tau_1 + b_2\tau_2 + id_3\tau_3.
\end{align}
That is, $\bm{b}$ and $\bm{d}$ are respectively as follows
\begin{subequations}
\begin{align}
  \bm{b} &= (b_1, b_2, 0)^T,\\
  \bm{d} &= (0, 0, d_3)^T.
\end{align}
\end{subequations}
Therefore, one of the conditionals (\ref{eq:condition})
is automatically satisfied by symmetry.
Since the system has two dimensions
and there are one conditionals,
$\bm{k}$ satisfying the conditionals
appears as one-dimensional objects (rings).
That is, these points are the SPERs.

\begin{figure}[tb]
  \begin{center}
    \includegraphics[scale=0.6, angle=90]{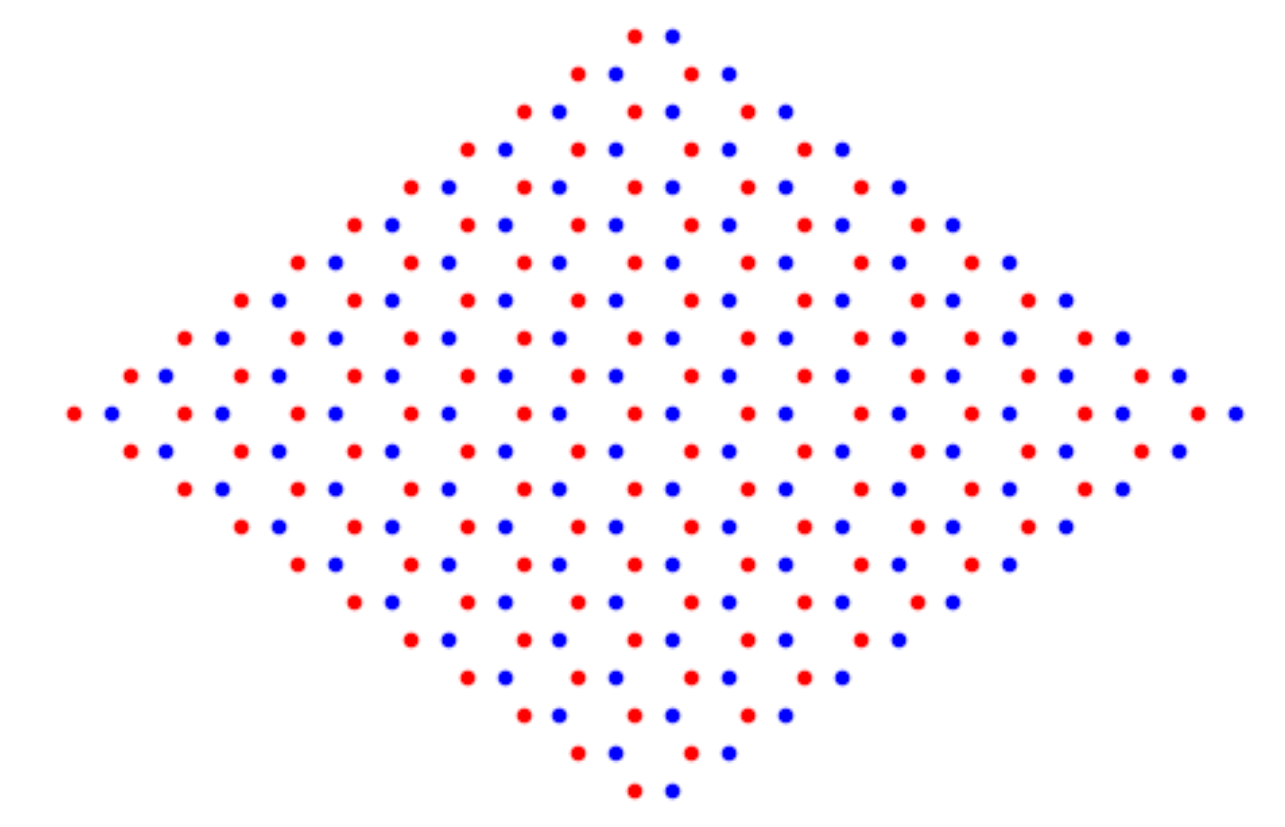}
    \caption{
    Sketch of the mechanical graphene
    used in the numerical calculations.
    There are 242 mass points in total.
    The unit cells are arranged
    so that it forms a rhombus.
    We consider the fixed boundary condition
    where the mass points of the boundaries are connected to
    the wall by springs.
    }
    \label{Fig:rhombus}
  \end{center}
\end{figure}%

\section{Properties of SPERs}
\label{App:02}
In fact, we can obtain the eigenvalue of $M(\bm{k})$ analytically.
When we consider the condition with the homogeneous friction
(in other words, matrix $\Gamma_0$ can be represented by
$-b\hat{1}$),
we can block diagonalize $M(\bm{k})$ using matrix $U(\bm{k})$
which diagonalize $D(\bm{k})$,
\begin{align}
  \left(
    \begin{array}{cc}
      U(\bm{k}) & 0\\
      0 & U(\bm{k})
    \end{array}
  \right)
  \left(
    \begin{array}{cc}
      0 & \hat{1}\\
      -D(\bm{k}) & -b\hat{1}
    \end{array}
  \right)
  \left(
    \begin{array}{cc}
      U^{\dagger}(\bm{k}) & 0\\
      0 & U^{\dagger}(\bm{k})
    \end{array}
  \right) \nonumber \\=
  \left(
    \begin{array}{cc}
      0 & \hat{1}\\
      -\hat{\omega}^2_0(\bm{k}) & -b\hat{1}
    \end{array}
  \right),
\end{align}
where $\hat{\omega}_0$ denotes the diagonal matrix
with eigenvalues of $D(\bm{k})$ on diagonal components.
The above equation can be rewritten as
\begin{align}
  \left(
    \begin{array}{cc}
      0 & \hat{1}\\
      -\hat{\omega}^2_0(\bm{k}) & -b\hat{1}
    \end{array}
  \right) =
  \bigoplus_n
  \left(
    \begin{array}{cc}
      0 & 1 \\
      -\omega_{0n}^2 & -b
    \end{array}
  \right),
\end{align}
where $n$ denotes the number of bands for $D(\bm{k})$.
From this equation, we can obtain the eigenvalues of $M(\bm{k})$,
\begin{align}
  \omega(\bm{k})
  = -\frac{b}{2}i \pm
  \sqrt{\omega_{0n}^2(\bm{k}) - \left(\frac{b}{2}\right)^2}.
\end{align}
This relation indicates that the SPERs emerge to satisfy the following equation,
\begin{align}
  \label{SPERcondition}
  \omega_{0n}(\bm{k}) = \frac{b}{2}.
\end{align}

\begin{figure}[tb]
  \begin{center}
    \includegraphics[scale=0.3]{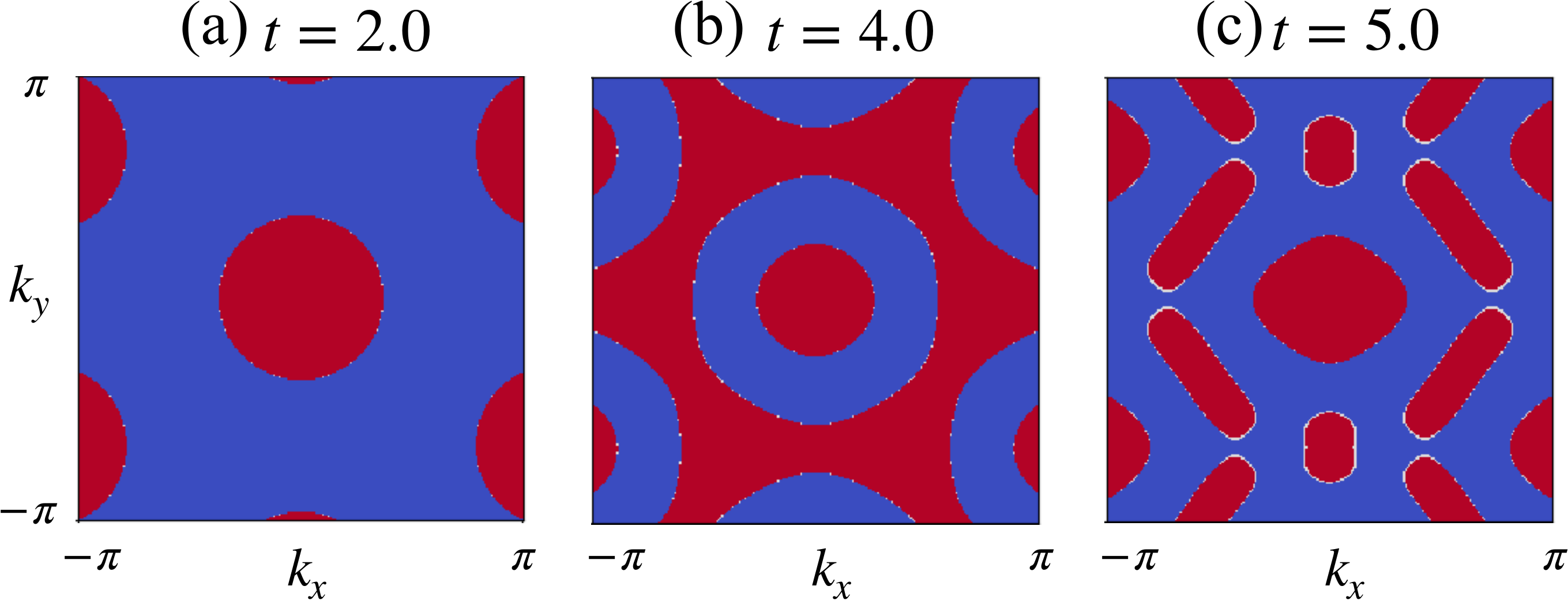}
    \caption{The dynamics of $u_{\bm{k}}$
    when the number of mass points is 50 with the fixed boundary
    (initial condition is $u_{\bm{k}}$ = 1 for all $\bm{k}$).
    (a), (b), and (c): Snapshots of time-evolution
    of $u_{\bm{k}}$ over all wavenumber
    (initial condition is $u_{\bm{k}}$ = 1 for all $\bm{k}$)
    at $t=2.0$, $4.0$, and $5.0$, respectively.
    }
    \label{Fig:50fixed}
  \end{center}
\end{figure}%

\section{System used in the numerical calculation}
\label{App:03}

In this section,
we show the details of the system that
we actually considered when performing the numerical calculations.
We also discuss how the dynamics is affected by the boundary
when the system is small.

\subsection{Cases not affected by boundaries}
Figure \ref{Fig:rhombus} shows
a sketch of the mechanical graphene used in the numerical calculations.
This system has boundaries, i.e.,
the mass points at the boundaries are connected to the wall by the spring.
The same results as for the periodic boundary condition are obtained
even when the system has the boundaries until
the oscillation is transmitted to the mass points at the boundaries.

\subsection{Cases affected by boundaries}
\label{App:06}
We show that in a system with a boundary,
the results for an infinite system cannot be reproduced
if the system is affected by reflected waves.

Figure~\ref{Fig:50fixed} shows the time-evolution of
$u_{\bm{k}}$ when the number of mass points is 50.
Up to a certain time,
the results are consistent for an infinite system,
but after the time, the results are clearly not consistent.
This is because the vibrations reach the edges
and are affected by the reflected waves.

\end{document}